\documentclass[a4paper,11pt]{article}
\usepackage{amsmath, amssymb, amsfonts,indentfirst,graphicx,color,anysize}
\usepackage{hyperref}
\marginsize{3cm}{3cm}{2cm}{2cm}
\title{
  \includegraphics[width=0.35\textwidth]{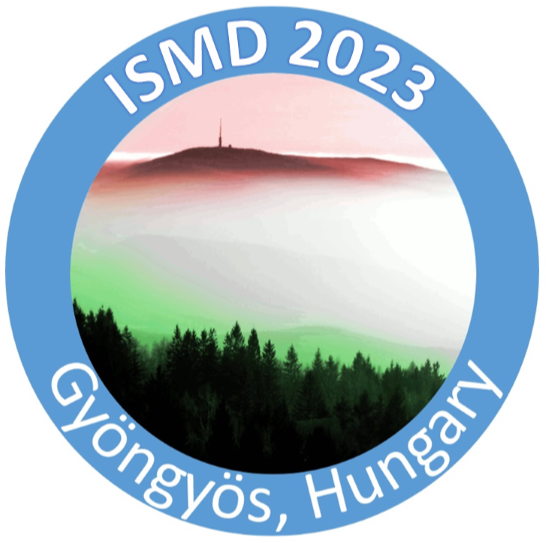}\\[1cm]
  \vskip -0.5cm
   \textbf{An Update on the Hypothetical X17 Particle }}

\author{{A.J. Krasznahorky$^1$, A. Krasznahorkay$^1$, M. Csatl\'os$^1$,
         J. Tim\'ar$^1$,} \\
        {M. Begala$^{1,2}$, A. Krak\'o$^{1,2}$, I. Rajta$^1$,
         I. Vajda$^1$, N.J. Sas$^{1,2}$}\\[1ex]
	$^1$Institute for Nuclear Research  (HUN-REN ATOMKI),\\
        P.O. Box 51, H-4001 Debrecen, Hungary\\
$^2$ University of Debrecen, Doctoral School of Physics, 4032 Debrecen,\\
  Egyetem t\'er 1, Hungary\\
}
\begin{document}

\maketitle

\begin{abstract} 

Recently, when examining the differential internal pair creation
coefficients of $^8$Be, $^4$He and $^{12}$C nuclei, we observed
peak-like anomalies in the angular correlation of the e$^+$e$^-$
pairs.
This was interpreted as the creation and immediate decay of an
intermediate bosonic particle with a mass of $m_{X}c^2\approx$ 17~MeV,
receiving the name X17 in subsequent publications.
Our results initiated a significant number of new experiments all over
the world to detect the X17 particle and determine its properties.  In
this paper we will give an overview of the experiments the results of
which are already published, and the ones closest to being published.
We will also introduce our latest results obtained for the X17
particle by investigating the e$^+$e$^-$ pair correlations in the
decay of the Giant Dipole Resonance (GDR) of $^{8}$Be.
  
\end{abstract}



\section{Introduction}

We published very challenging experimental results in 2016 \cite{kr16}
indicating the electron-positron (e$^+$e$^-$) decay of a hypothetical
new light particle.  The e$^+$e$^-$ angular correlations, measured with a newly built spectrometer \cite{gu16} for the
17.6~MeV and 18.15~MeV transitions in $^8$Be were studied and an
anomalous angular correlation was observed for the 18.15~MeV
transition \cite{kr16}. This was interpreted as the creation and immediate decay
of an intermediate bosonic particle with a mass of
$m_{X}c^2$=16.70$\pm$0.35(stat)$\pm$0.5(sys)~MeV, receiving the name
X17 in subsequent publications.

Our results were first interpreted theoretically with a new vector gauge boson
by Feng and co-workers \cite{fe16,fe17,fe20}, which would mediate a
fifth fundamental force
with some coupling to standard model (SM) particles. The possible relation of
the X17 boson to the dark matter problem triggered an enormous interest in the
wider physics community \cite{ins} and resulted in also many other
interpretations,
the complete survey of which is beyond the scope of this paper.

We also observed a similar anomaly in $^4$He \cite{kr21}. It could be
described by the creation and subsequent decay of a light particle
during the proton capture process on $^3$H to the ground state of
$^{4}$He. The derived mass of the particle
($m_{X}c^2 = 16.94 \pm 0.12(stat)\pm 0.21(sys)$~MeV)
agreed well with that of the proposed X17 particle.

Recently, we have studied the E1  decay of the 17.2~MeV
J$^\pi$ = 1$^-$ resonance to the ground state in $^{12}$C \cite{kr22}.
The angular
correlation of the e$^+$e$^-$ pairs produced in the
$^{11}$B(p,$\gamma$)$^{12}$C reaction were studied at five different
proton energies around the resonance. The gross features of the
angular correlations can be described well by the Internal Pair
Creation (IPC) process following the E1 decay of the $1^-$ resonance.
However, on top of the smooth, monotonic distribution, we observed
significant peak-like anomalous excesses around 155-160$^\circ$ at four
different beam energies. The e$^+$e$^-$ excess could be well-described by
the creation and subsequent decay of the X17 particle.  The invariant
mass of the particle was derived to be ($m_{X}c^2 = 17.03 \pm 0.11(stat.)\pm 0.20$(syst.)~MeV),
in good agreement with our previously published values.

In conclusion, the well defined excitation energy of the nucleus
after the proton capture is used to create a new particle, and the
rest gives kinetic energy for the created particle.
The larger the kinetic energy, the smaller the opening angle between the e$^+$e$^-$
pairs, according to the formulas derived for the two particle decay of
a moving particle. This provides strong kinematic evidence that  all
results were caused by the same particle as concluded by Feng et al.
\cite{fe17}.

However, despite the consistency of our observations, more
experimental data are needed to understand the nature of this
anomaly. For this reason, many experiments all over the world are in
progress to look for such a particle in different channels. Many of
these experiments have already put constraints on the coupling of this
hypothetical particle to ordinary matter. Others are still in the
development phase, but hopefully they will soon contribute to a deeper
understanding of this phenomenon as concluded by the community report
of the Frascati conference \cite{al23} organized in 2022.

The aim of this paper is to give an overview of the experiments the
results of which are already published,
and the ones which are closest to being published.

We also discuss our latest
results obtained for the X17 particle during the investigation of the
Giant Dipole Resonance (GDR) decay.

\section{Overview of the most important experiments searching for
  the X17 particle}

Every new elementary particle, especially bosons, could be associated with a new
force or at least with a new, unknown or unexpected aspect of one of
the known forces.  More in general, the possible existence of a new
particle is of paramount importance to particle physics and
cosmology.  Therefore, our experiments at ATOMKI
initiated a significant number of new experiments to detect the X17
particle and determine its properties.

\subsection{Confirming the $^8$Be anomaly with a two-arm electron
  positron pair spectrometer in Hanoi}

The newest experimental searches for the X17 particle performed by Tran The
Anh et al., at the Hanoi University of Sciences, Vietnam confirmed  the
presence of the X17 anomaly in $^8$Be \cite{anh24}.

We successfully built a two-arm e$^+$e$^-$ spectrometer in
collaboration with the Hanoi University of Science. The spectrometer
was tested and calibrated using the 17.6 MeV M1 transition excited in
the $^7$Li(p,e$^+$e$^-$)$^8$Be reaction. They have obtained a nice
agreement between the experimentally determined acceptance of the
spectrometer and the one coming from their simulation.

They measured the angular correlation of the e$^+$e$^-$ pairs at E$_p$ =
1225 keV, which is the off-resonance region above the E$_x$ = 18.15 MeV
state.  In agreement with the ATOMKI results they have observed a significant anomaly ($\geq 4\sigma$)
around 135 degree for this state, but did not observe any anomaly for the 17.6 MeV state \cite{kr16}.

\subsection{Observation of structures at ~17 and ~38 MeV/c$^2$
in the $\gamma\gamma$ invariant mass spectrum in d+Cu collisions}

 Abramyan et al.,
at the Joint Institute for Nuclear Research, Dubna, Russia reported evidence
on the observation of X17 in the $\gamma\gamma$ invariant mass
spectra in  d + Cu collisions at p$_{lab}$ of a few GeV/c per
nucleon \cite{abra24}.

They observed enhanced structures at
invariant masses of about 17  and 38 MeV/c$^2$
in the reaction d + Cu $\rightarrow$ $\gamma + \gamma$ + X at a
momentum of 3.83 GeV/c per nucleon.
The $\gamma$-rays were detected by 32 lead glass scintillation spectrometers,
which were placed 300 cm from the target.
The significance of the peak observed at 17 MeV/c$^2$ was better than 6$\sigma$.
The results of
testing of the observed signals, compared with the results of
the Monte Carlo simulation support the conclusion
that the observed signals are the consequence of
detection of the decay of particles with masses of about 17 and 38 MeV/c$^2$
into a pair of photons.
The presented evidence of both X17 and E38, suggests that there
might be several new particles below the mass of the $\pi^0$ particle.

The experiment was repeated also for  e$^+$e$^-$ pairs and observed a
significant peak also in their invariant mass spectrum at
17 MeV/c$^2$ \cite{ab24}.

\subsection{The MEG II (Muon Electron Gamma) experiment}

To verify the existence of the X17 particle, experiments using the
$^7$Li(p,$\gamma$)$^8$Be nuclear reaction were carried out  at the
Paul Scherrer Institute with the MEG II (Muon Electron Gamma)
superconducting solenoid spectrometer \cite{an24}.

Inside the solenoid a special drift chamber was used to
determine the four-momenta of the
electrons and positrons moving on spiral paths emitted from the $^7$Li(p, e$^+$e$^-)^8$Be reaction, on the Li target placed in the center of the solenoid.  Feasibility studies,
performed with a complete detector simulation and including realistic
background models, suggest that a 5$\sigma$ sensitivity could be
reached with this setup. The analysis of the results of the experiment performed
in 2023 is already in its last phase. Their results are expected
to become public soon. Their plans and the description
of their spectrometer have been presented at several
conferences so far, and a PhD thesis is being prepared from them.

Recently,  Papa \cite{an24} presented their latest experimental results at the
ICHEP 2024 conference in Prague.  She reported on the status of this
search with the MEG II apparatus, presenting the collected data, the
analysis strategy, and the current results.

They did not use any analysing magnet after the Cockroft-Walton
accelerator and in this way the beam contained a reasonable amount
($\approx$25\%) of H$_2$ molecules, making the measurement on resonance
more complicated.  Moreover, they used a $\approx 7 \mu$ m thick LiPON
target in which the beam is fully stopped and both the 440~keV and
1030~keV resonances would be excited simultaneously with an unfavorable
intensity ratio of 79.2\% and 20.8\% from the 440~keV and 1030~keV
resonances, respectively. In this way they could not study the
resonances individually, which could have resulted in a reduced
background. Having a low background is very important,
especially for the 1030~keV resonance, from which decay we observed
the anomaly earlier.  Even worse, they took care of the target
cooling with a massive target frame and support rod, which produced
a large background from external pair creation.

Within the above conditions the expected signal to background ratio
of their results
will be much lower compared to the one we observed in ATOMKI before,
which may prevent the detection of the anomaly.

She concluded that the analysis is well advanced and ready to report
the results soon for the background regions, but not for the signal
region yet. A new X17 data collection, fully exploiting the 1030~keV
resonance is foreseen during the first part of 2025 (Physics Run 2025).

\subsection{The Mu3e  experiment, using the MEG II setup}

Another channel that will be investigated with the Mu3e experiment is
$\mu^+ \rightarrow e^+ X$ in which $X$ is an axion-like particle from a
broken flavour symmetry like a familon or majoron that leaves the
detector unseen \cite{ann23}. For this purpose, the Mu3e data acquisition is
adapted to accommodate online histogramming of track fit results such
as momenta and emission angles of events with single positrons on the
event filter farm. Current limits on $\mu^+ \rightarrow e^+ X$ are set
by Jodidio et al. al., B($\mu^+ \rightarrow e^+ X$)$<2.6\times
10^{-6}$ at 90 \% CL for massless X  \cite{ann23}, and by the TWIST
collaboration at $< 9\times 10^{-6}$ for X with masses between 13~MeV
and 80~MeV.  The sensitivity of the Mu3e experiment in phase I exceeds
the limits set by TWIST by two orders of magnitude in a large range of
X masses and will be further improved in phase II due to a twenty
times larger number of observed muon decays and an enhanced detector
performance.

\subsection{The PADME (Positron Annihilation for Dark Matter Experiment}

Experiments using positron beams impinging on fixed targets offer
unique capabilities for probing new light dark particles weakly
coupled to e$^+$e$^-$ pairs, that can be resonantly produced from
positron annihilation on target atomic electrons.

A study of the resonant production of the X17 with a positron beam
started at Laboratori Nazionali di Frascati (LNF) in 2022 \cite{al23}.
The year 2023 was dedicated to the analysis of the data collected in
Run-III for the X17 campaign and the latest results were published at
the ICHEP 2024 conference by Venelin Kozhuharov \cite{ko24}.

Taking advantage of the unique
opportunity to have positrons in the energy range 250-450~MeV, PADME
is in an ideal position to produce the X17 state in a resonant mode
and subsequently detect it via its decay to an e$^+$e$^-$ pair. The
PADME Run-III (from October to December 2022) consisted of an energy
scan of the X17 mass region.  The main background to the X17
$\rightarrow$ e$^+$e$^-$ signal is the elastic (Bhabha)
electron-positron scattering. While the t-channel is peaked at high
energies for the scattered positron, the s-channel has the same signal
kinematics. The expected peak to background ratio (assuming a vector
particle) is only about 0.6-2.0\% for the allowed coupling constant
region, making it very challenging to find this resonance.  The RUN III
analysis is in its final track. The number of positrons on target is
determined with various cross-calibration procedures with an uncertainty
$<$ 1\%.  The signal acceptance and background estimation are reported to be under
control with systematics of O(1\%). Public results from the experiment
are expected by the end of this year.

They are planning to have another run (RUN IV) in early 2025 with an
upgraded detector.  A major improvement to PADME's setup for RUN IV
includes precise e$^+$e$^-$/$\gamma\gamma$ discrimination with a
Micromegas tracker.  They want to have 4 times higher statistics per
scan point with higher beam intensity by a factor of 2 and with less
scan points due to the widening of the X17 lineshape because of the
electronic motion. Such improvements will hopefully allow probing the full
unexplored region for the X17 particle.

\subsection{Searching for the X17 particle with a highly efficient
  e$^+$e$^-$ spectrometer in Montreal}
At the Montreal Tandem accelerator, an experiment is being set up to
measure the electron-positron pairs from the decay of the X17 particle,
using the same $^{7}$Li($p$,$\gamma$)$^{8}$Be nuclear reaction studied
in ATOMKI \cite{kr16}, for an independent observation of the X17 particle
\cite{az22}.
They are using long multiwire proportional chamber and scintillator
bars, surrounding a target. The beam travels along the symmetry axis of the
spectrometer with the target located in the middle. The most
important feature of their spectrometer is its nearly 4$\pi$ solid
angle coverage.  The $^7$LiF target will be mounted on
an Al foil and water-cooled in a thin carbon fiber section of the
beamline.  Assuming the ATOMKI evaluation of the electron-pair
production rate from X17, Geant4 simulation predicts the observation of a
clear signal after about 2 weeks of data taking with a 2~$\mu$A proton
beam. Presently they are still in the process of setting up the
apparatus.

\subsection{The New JEDI (Judicious Experiments for Dark sectors Investigations) experiment at GANIL}

At GANIL France they plan to develop a long-term research program in
the MeV terra incognita energy range at the new SPIRAL2 facility
(Caen, France), that will deliver unique high-intensity beams of
light and heavy-ions, and neutrons in Europe.

For measuring the electron-positron angular correlations a set of
Double-sided Silicon Strip Detectors (DSSDs) of the New JEDI
(Judicious Experiments for Dark sectors Investigations) setup will provide
energy losses and angles of the detected electrons and positrons
\cite{ba24}.

In addition, sets of plastic detectors will be used to measure the
residual energy of electrons and positrons, and to veto
external background events. The detection system is coupled to new
generation NUMEXO2 digitizers. A specific charge integration firmware
and a coincidence cross-check mode have been developed for the
project. It is worth noting that the New JEDI detection system appears
to be close to the new version of the spectrometer developed in our
ATOMKI group. With the geometry chosen at GANIL it is more focused on the
detection of X17-like events.  To start off, they plan to
populate ``excited'' non-resonant states in $^3$He around 18~MeV using the
high-power pulsed proton beam of the LINAC impinging onto a thin CD$_2$
target.

\subsection{A new experimental setup at LNL, Legnaro}

The building block of the setup discussed in Ref.\cite{al23} at  p.20
is a plastic scintillator $\Delta$E-E
telescope composed of three detector layers.  The E stage is built
using a 5$\times$5$\times$10 cm EJ200 scintillator read out by a Silicon
PhotoMultiplier (SiPM).  The $\Delta$E stage has two sub-layers: each one is
made of 10 EJ200 strips of dimensions 0.5$\times$0.2$\times$5 cm, read out by
an array of 2$\times$2 mm SiPMs.  By placing the bars of the second
layer orthogonally with respect to the first one, a grid is obtained,
allowing the measurement of both coordinates of a particle's entry positions into
the telescope.

The telescopes are organized in groups of 4, forming a clover held by a
plastic cage. The clovers will be placed at 15~cm from the center of
the target, with the $\Delta$E layer facing it. The project plans to produce
and use a minimum set of five clovers, placed at different angular
positions to get as uniform of an acceptance as possible. The device can be
operated in vacuum, making it possible to minimize the amount of matter seen by
the particles before reaching the first detection layer.  The project
is still under construction.

\subsection{X17 experiments of the nTOF (neutron Time of Flight)
  collaboration at CERN}

An Italian group is engaged to carry out a first series of measurements at
nTOF, where the excited levels of $^4$He can be populated via the
conjugated $^3$He(n,e$^-$e$^+$)$^4$He  reaction using the
spallation neutron beam EAR2 at CERN \cite{ge23,gu24}. Their preliminary theoretical
studies indicate a much higher cross section when neutron
transitions are observed with respect to the ones of protons
(protophobic scenario). This approach has two relevant advantages: (i)
for the first time the X17's existence would be investigated through neutron
induced reactions exploiting the unique properties of the EAR2 beam
[6] and (ii) the experimental setup is completely different from
to the one used by our ATOMKI group. 
To measure the kinematics of the created particles, reaching a high level
of particle identification in a wide energy range and to optimize the
signal-to-noise ratio,  they propose a
detection setup based on two TPC trackers of rectangular shape
(50x50x5~cm) placed at both side of the target, backed by 50x10x10~cm long EJ200
slabs. Part of this setup was successfully tested recently in ATOMKI,
Debrecen. The goal of that experiment was to establish the capability of
the demonstrator of a new e$^+$e$^-$ spectrometer to measure the 4-momenta
of IPC e$^+$e$^-$ pairs generated in the $^7$Li(p,e$^-$e$^+$)$^8$Be reaction.
They may
still need a few years to get experimental results at CERN.

\subsection{The search for X17 at the Czech Technical University in Prague}

The Chech group proposes finalizing an existing
spectrometer composed of six small TPCs equipped with multiwire
proportional counters (MWPC) at their entrance windows, and its
upgrade with an inner tracker based on Tpx3 detectors. It would consist of
a cylindrical vessel, divided into sextants, separated by strong
permanent magnets. The spectrometer could easily be installed into or removed
from the beam line \cite{hu23}.

Each sextant of the spectrometer could be operated as a separate
detector, and be composed of three sensitive layers. The first
layer would be a Tpx3 detector with a very thin Si sensor (50~$\mu$m) and its
ASIC. Its 55~$\mu$m spatial resolution would provide an excellent angular
resolution and a very precise determination of the source
vertex of the detected particles. The second layer is planned to be inside the gas volume and consist of a
MWPC that would provide some redundancy in the angular determination of the
system. It would also be used to correct the energy as a function of
the scattering angle of the particles in the Tpx3 detectors and the vacuum
tube wall.
The momentum of the particles would finally be measured by the TPCs in a
toroidal magnetic field.

\subsection{Particle and Nuclear Physics at the MeV scale in Australia}

An international group intends to employ the Pelletron accelerator in Melbourne to
initiate nuclear reactions of the kind: p+ZX$\rightarrow$ (Z+1)Y + (e$^+$e$^-$) and to
build a low mass, high precision Time Projection Chamber (TPC), with
world-first capabilities \cite{se23}.  The invariant mass resolution of the TPC for
the e$^+$e$^-$ final state is expected to be 0.1~MeV. This would provide a
substantially more sensitive search for anomalous e$^+$e$^-$ production
than any other experiment and be 200 times more sensitive than experiments done at
ATOMKI. Accordingly, they will either observe the ATOMKI anomaly on the
Pelletron or exclude it at a very high significance. Following this they
propose a program to search for anomalous e$^+$e$^-$ production with
world-leading sensitivity in the 5-25~MeV mass region. In addition,
the very large acceptance, and excellent angular and energy resolution
of the TPC would enable qualitatively more sensitive investigations of
Nuclear Internal Pair Conversion decays. This capability would allow for a
range of novel Nuclear Physics investigations.

\subsection{The PRad experiment at JLab}

A new electron scattering experiment (E12-21-003)
has been approved at Jefferson Lab
to verify and
understand the nature of hidden sector particles, with particular
emphasis on the X17 particle \cite{du23}.
The proposed direct detection experiment will use a
magnetic-spectrometer-free setup (the PRad apparatus) to detect all
three final state particles in the visible decay of a hidden sector
particle for an effective control of the background and will cover the
proposed mass range in a single setup. The use of the
well-demonstrated PRad setup allows for an essentially ready-to-run
and uniquely cost-effective search for hidden sector particles in the
3-60~MeV mass range with a sensitivity of 8.9$\times$10$^{-8}$ to 5.8$\times$10$^{-9}$ in
the square of the kinetic mixing interaction constant between hidden
and visible sectors.

Other experiments where it will be possible to search
for the X17 are FASER  \cite{fa18}  at CERN, DarkLight  \cite{co17}  and
HPS  \cite{mo13}  at JLAB, VEPP-3  \cite{wo18} at Novosibirsk and the
MAGIX and DarkMESA experiments foreseen at the MESA
accelerator complex at Mainz  \cite{hu17}.
In addition, searches in charmed meson and J/$\Psi$ decays have
been proposed  \cite{ca21,ba20}, that can be explored at Belle II
(SuperKekB), BESIII (BEPCII) and LHCb (CERN).

\section{Observation of the X17 anomaly in the decay of the Giant Dipole Resonance of $^8$Be}

Pleased with the successful experiments performed with the double-arm
e$^+$e$^-$ spectrometer in Hanoi \cite{anh24}, we set up an
experiment with a similar spectrometer at ATOMKI.
The  simpler geometry of such a spectrometer, and the smooth acceptance curve
as a function of the relative angle of the e$^+$e$^-$ pairs
avoid non-trivial possible artefacts, which might be connected to the
spectrometer itself \cite{al21}.

With such a new spectrometer, we studied the X17 creation and the
e$^+$e$^-$ pair emission from the decay of the Giant Dipole Resonance
(GDR) \cite{fi76,sn86,ha01} excitations of $^8$Be.

\subsection{The e$^+$e$^-$ spectrometer}

\begin{figure}[htb]
  \begin{center}
    \includegraphics[scale=0.5]{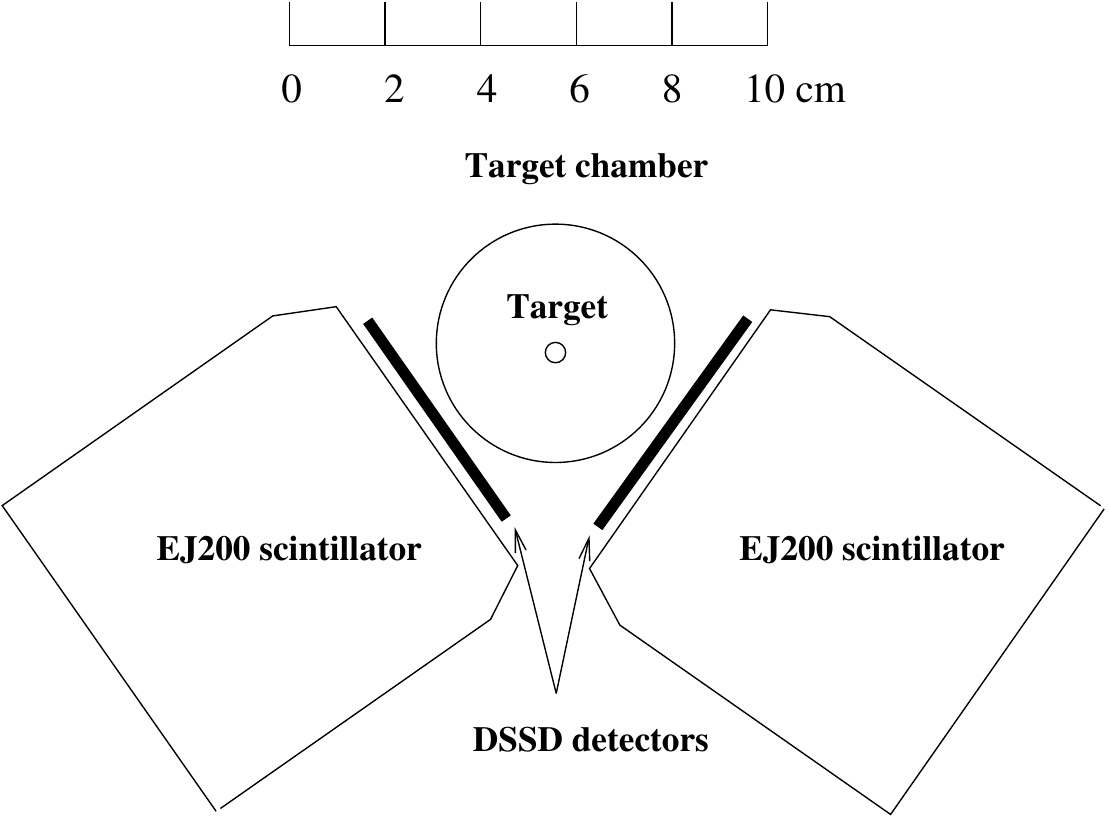}
  \end{center}
  \caption{Schematic diagram of the e$^+$e$^-$ spectrometer.}
  \label{fig:setup}
\end{figure}

In the present experiment two detector telescopes consisting of
Double-sided Silicon Strip Detectors (DSSD) and plastic scintillators
were used, placed at an angle of 110$^\circ$ with respect to each
other as shown in Fig~\ref{fig:setup}.

The diameter of the carbon fiber tube of the target chamber has
been reduced to 48~mm to allow a closer placement of the telescopes to
the target.  This way we could cover a similar solid angle as our
previous setups, which had more telescopes.

The two detector telescopes were placed at azimuthal angles
-35$^\circ$ and -145$^\circ$ with respect to the horizontal 0$^\circ$ angle. This meant
that cosmic rays, predominantly arriving vertically, would have a very
low chance of hitting both telescopes at the same time.

The DSSD's were used to measure the energy loss of the e$^+$e$^-$
particles and their directions. They consist of 16 sensitive strips on
the junction side and 16 orthogonal strips on the ohmic side. Their
element pitch is 3.01~mm for a total coverage of 49.5$\times$49.5~mm$^2$.

The dimensions of the EJ200 plastic scintillators were
82$\times$86$\times$ 80 mm$^3$,
each connected to Hamamatsu 10233-100 PMT assemblies.
In order to place the detectors as close as possible to the target,
their front sides were trapesoidally shaped.

During our previous measurements, we examined the decay of excited
states for which the emission of neutrons was not an allowed
process. In this way, we did not have to worry about radiation damage
to our DSSD detectors.  However, the GDR examined in our present work
decays primarily by the emission of neutrons, which can destroy the
DSSD detectors in our setup. We estimate 9-10 orders of magnitude more
neutrons produced, than e$^+$e$^-$ decays of the X17 particle.

To actively shield against the effect of cosmic radiation, we used 13 plastic
scintillators measuring 100$\times$4.5$\times$1~cm$^3$, which were placed above the
spectrometer. The signals from these  detectors were fed into CFD
discriminators and then into TDCs. If one of these detectors fired, that event
was omitted from the offline data analysis. Since the efficiency of the
cosmic shielding was only 50\%, after the measurements, we
also performed a cosmic background measurement for the same amount of
time, as we had with beam on target. The cosmic events  collected in this way
were subtracted from our  spectra.

\section{Experiments}

The experiments were performed in Debrecen (Hungary) at the 2~MV
Tandetron accelerator of ATOMKI, with a proton beam energy of E$_p$=
4.0~MeV. Owing to the rather large width of the GDR ($\Gamma$~=~5.3~MeV
\cite{fi76}), a 1~mg/cm$^2$ thick $^{7}$Li$_2$O target was used in
order to maximize the yield of the e$^+$e$^-$ pairs. The target was
evaporated onto a 10 $\mu$m thick Ta foil. The average energy loss of
the protons in the target was $\approx$100~keV.

$\gamma$ radiation was detected by a 3''x3'' LaBr$_3$ detector,
monitoring also for any potential target losses. The detector was placed
at a distance of 25~cm from the target at an angle of 30$^\circ$ to
the beam direction.

\begin{figure}[htb]
  \begin{center}
    \includegraphics[scale=0.4]{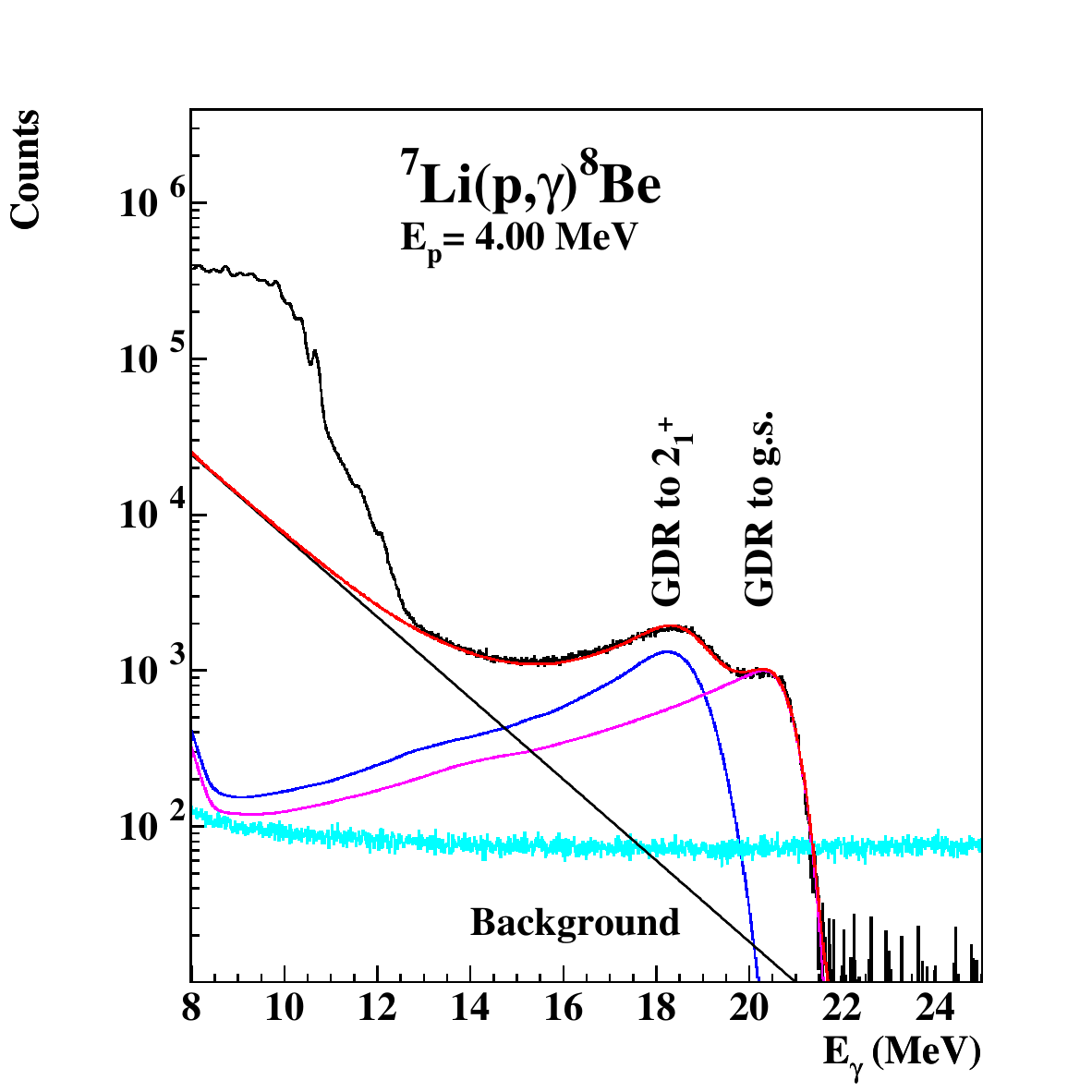}
  \end{center}
  \caption{Tipical $\gamma$-ray spectrum measured for the
    $^{7}$Li($p$,$\gamma$)$^{8}$Be nuclear reaction at $E_p$~=~4.0~MeV}
  \label{fig:gamma}
\end{figure}

A typical $\gamma$ energy spectrum is shown as a black histogram in 
Fig~\ref{fig:gamma}. The figure clearly shows the transitions from the
decay of GDR to the ground and first excited states in $^{8}$Be. The
cosmic ray background is shown in cyan and was already subtracted from
the signal distribution. It was found to be low, and reasonably uniform.

The simulated spectra for the ground-state and first excitation state
transitions, including their natural widths and the resolution of the
detector is shown in red and blue, respectively. The remaining
background, coming mostly from neutron capture on the surrounding
materials was estimated by an exponential curve and used to fit the
$\gamma$-spectrum between 14 and 22~MeV together with the simulated
response curves. The result of the fit is shown in red in  Fig~\ref{fig:gamma}.
The intensity ratio of the peaks, obtained from the fit is: I(GDR$\rightarrow $
g.s.)/I(GDR$\rightarrow 2_1^+$)~=~0.88$\pm$0.09 at E$_p$~=~4.0~MeV and $\theta$~=~30$^\circ$.
It is consistent with the results of Fischer et al. \cite{fi76}.

At the proton energy of E$_p$~=~4.0~MeV, the (p,n) reaction channel is
open (E$_{thr}$~=~1.88~MeV), generating neutrons and low-energy
$\gamma$ rays with a large cross section.  (Other reaction channels
are also open, but their cross sections are much smaller and their
influence on our experiment is much weaker.)  The maximum neutron
energy (E$_n$~=~1.6~MeV) induces only a 300~keV electron equivalent
signal in the plastic scintillator due to the quenching effect. Such a
small signal fell well below the CFD thresholds that we used.
The low-energy neutrons did not produce any measurable signal in the
DSSD detectors either. The maximum energy that can be transferred
in elastic scattering on Si atoms is only $\approx$50~keV, which is
below the detection threshold.

\subsection{Calibration of the acceptance of the spectrometer}
\label{sec:acceptance}

The acceptance calibration of the whole e$^+$e$^-$ coincidence pair
spectrometer was performed in a similar way as described in
Ref. \cite{kr21}.  It was crucial for the precise angular correlation
measurements to measure and understand the response of the whole
detector system to isotropic e$^+$e$^-$ pairs as a function of the
correlation/opening angle.

The detectors measure continuous e$^+$e$^-$ spectra and the sum of the
energies are constructed offline. Due to the energy loss in the wall
of the vacuum chamber and in the DSSD detectors, as well as the finite
thresholds of the discriminators (CFD), the low-energy part of the
spectrum is always cut out. Since we measure e$^+$e$^-$ coincidences,
such a low energy cut also means a high energy cut for the particles
detected in coincidence.  Thresholds were set to have similar
efficiencies in the different telescopes. After a proper energy
calibration of the telescopes, this was done in the analysis software.
The response curve was found to depend primarily on the geometrical
arrangement of the two detector telescopes.

Beside the e$^+$e$^-$ coincidences, down-scaled single events were
also collected during the whole run of the experiment for making
acceptance/efficiency calibrations. An event mixing method explained
in Ref.~\cite{kr21} was used to experimentally determine the relative
response of the spectrometer as a function of the correlation angle by
using the single telescope triggered events.  Uncorrelated lepton
pairs were generated from subsequent single events and their
correlation angle was calculated as for the coincident events. The
resulting angular correlation for the uncorrelated events gave us the
experimental response curve.

\begin{figure}[htb]
  \begin{center}
    \includegraphics[scale=0.4]{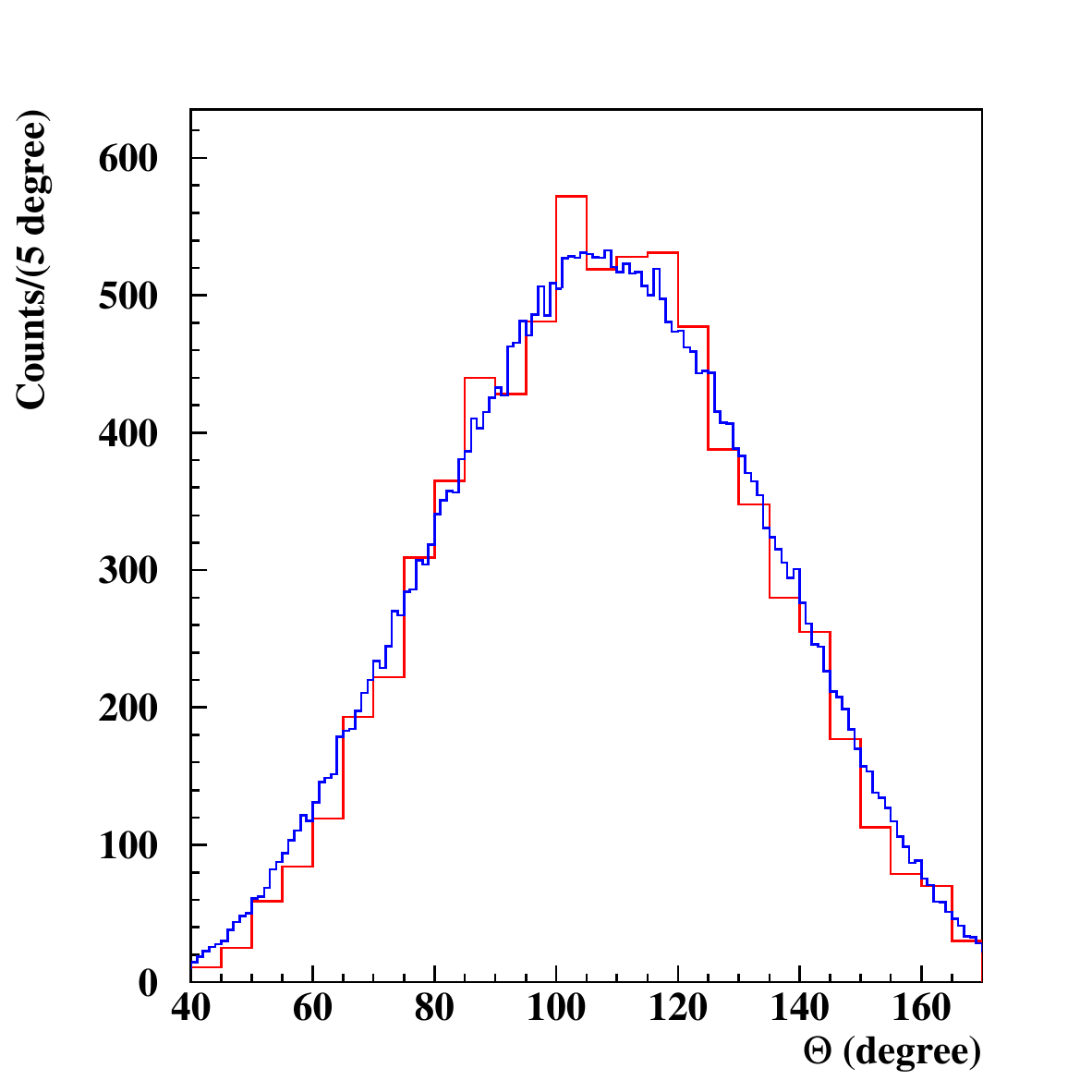}
  \end{center}
  \caption{Experimental acceptance of the spectrometer as a function of
    correlation angle ($\theta$) for consecutive, uncorrelated  e$^+$e$^-$
    pairs (red line histogram) compared with the results of the MC
    simulations (blue line histogram) as explained in the
    text.}
  \label{fig:acceptance}
\end{figure}

Reasonably good agreement was obtained between the efficiencies
measured in data and the Monte Carlo simulations,
as presented in Fig.~\ref{fig:acceptance}. The
average difference between the two is within $\approx 7.0$\% in the
70$^\circ$~-~170$^\circ$ range.

Due to the very tight geometry, the DSSD position data and therefore
the e$^+$e$^-$ angular distribution experiences an enhanced dependence
on the beam spot size and position. According to previous measurements
and MC simulations of the present setup we could take this effect
into account properly.

\subsection{Results for the angular correlation of the e$^+$e$^-$ pairs}

\begin{figure}[htb]
\begin{center}
\includegraphics[scale=0.4]{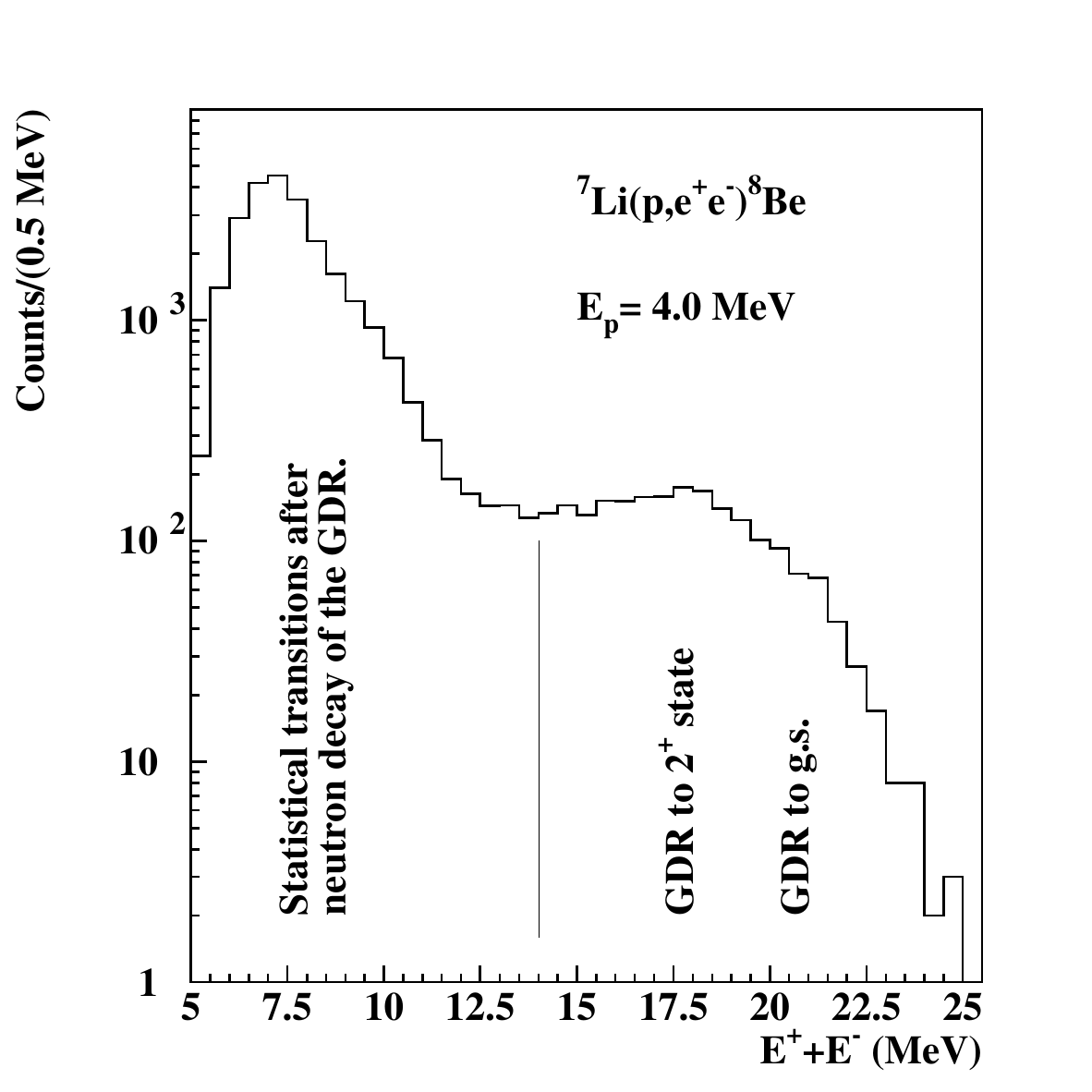}
\end{center}
\caption{Total energy spectrum of the e$^+$e$^-$ pairs from the
  $^{7}$Li(p,e$^+$e$^-$)$^{8}$Be nuclear reaction.}
\label{Fig:sume}
\end{figure}

The energy sum spectrum of the two telescopes are shown in
Fig.~\ref{Fig:sume}. The angular correlation spectra of the e$^+$e$^-$
pairs for the different energy sum regions were then obtained for
symmetric $-0.5\leq\epsilon \leq 0.5$ pairs. Here the energy
asymmetry parameter, $\epsilon$, is defined as $\epsilon=(E_1-E_2)/(E_1
+ E_2)$, where $E_1$ and $E_2$ denote the kinetic energies of the
leptons measured in telescopes 1 and 2, respectively.

\begin{figure}[htb]
    \begin{center}
\includegraphics[scale=0.35]{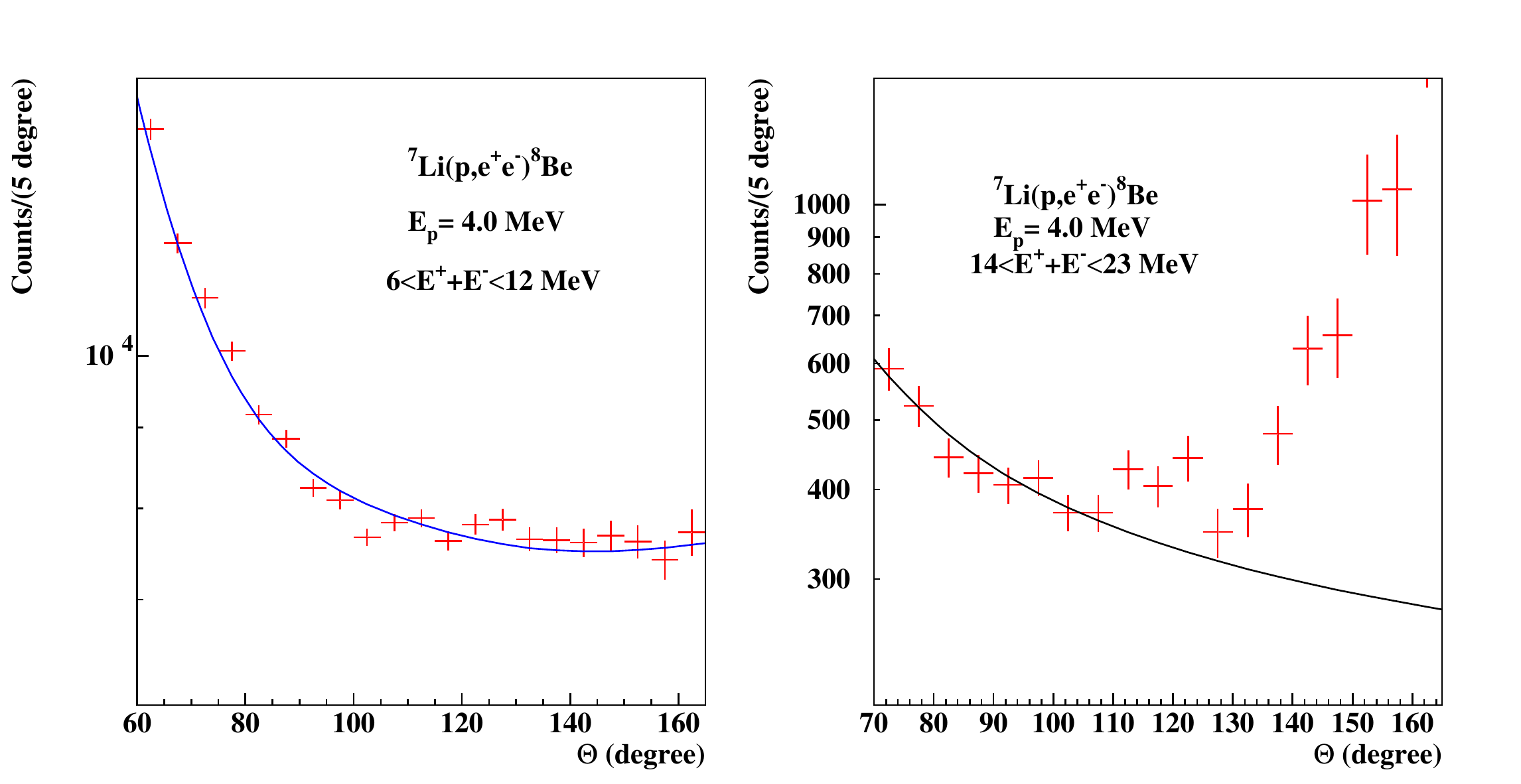}
    \end{center}
\caption{Left side: Experimental angular correlations of the e$^+$e$^-$
  pairs measured in the $^{7}$B(p,e$^+$e$^-$)$^{8}$Be reaction at
  E$_p$~=~4.0~MeV for low-energy ($E_1 + E_2 \leq$~14~MeV)
  transitions. Right side: the same for the high energy (GDR) region
  (14~$< E_1 + E_2 \leq$~25~MeV).}
     \label{Fig:ang-low-high}
\end{figure}

The angular correlation gated by the low energy-sum region (below 14~MeV),
as marked in Fig.~\ref{Fig:sume}, is shown in the left side of
Fig.~\ref{Fig:ang-low-high}. The measured counts were corrected for the
acceptance obtained as described in Section \ref{sec:acceptance}.  It
is a smooth distribution without showing any anomalies. It could be
described by assuming E1 + M1 multipolarities for the IPC process and
a constant distribution, which may originate from cascade transitions
of the statistical $\gamma$ decay of the GDR appearing in real
coincidence. In such a case, the lepton pair may come from different
transitions, and thus their angles are uncorrelated.  This smooth
curve reassured us that we were able to accurately determine the
efficiency of the spectrometer.
The angular correlation of the e$^+$e$^-$ pairs gated by the GDR energy
region (above 14~MeV), as marked in Fig.~\ref{Fig:sume}, is shown in the right
side of 
Fig.~\ref{Fig:ang-low-high}.

The experimental data corrected for the acceptance of the spectrometer
is shown as red dots with error bars. The simulated angular
correlation for the E1 internal pair creation is indicated as a black
curve. Significant deviations were observed. First of all, a peak-like
deviation at 120$^\circ$, but also an even stronger deviation at
larger angles.

The measured angular correlation was fitted from 70$^\circ$ to 160$^\circ$
with the sum of simulated E1, M1 and X17 contributions
calculated for both the GDR to ground state and for the GDR to 2$_1^+$
state transitions. The simulations concerning the decay of the X17
boson in the transition to the ground state of $^8$Be were carried out
in the same way as we did before \cite{kr16,kr21,kr22} and could
describe the anomaly appearing at around 120$^\circ$.

However, based on Fig~\ref{fig:gamma} and previous
measurements \cite{fi76}, the $\gamma$-decay of GDR to the first
excited state is stronger than its decay to the ground state.
According to that, we assumed that the X17 particle was created 
in the decay of GDR to both the ground state and to the first excited
state.  Based on the energy of that transition (17.5~MeV), we would
expect a peak around 150$^\circ$. However, the first excited state is
very broad ($\Gamma$~=~1.5~MeV), so the shape of the expected anomaly is
significantly distorted.  The simulations were then performed as a
function of the X17 mass from 10~MeV/c$^2$ to 18~MeV/c$^2$ for both
transitions.

To derive the invariant mass of the decaying particle, we carried out
a fitting procedure for both the mass value and the amplitude of the
observed peaks. The fit was performed with RooFit
\cite{Verkerke:2003ir} in a similar way as we described before
\cite{kr21,kr22}.

\begin{figure}[htb]
    \begin{center}
\includegraphics[scale=0.4]{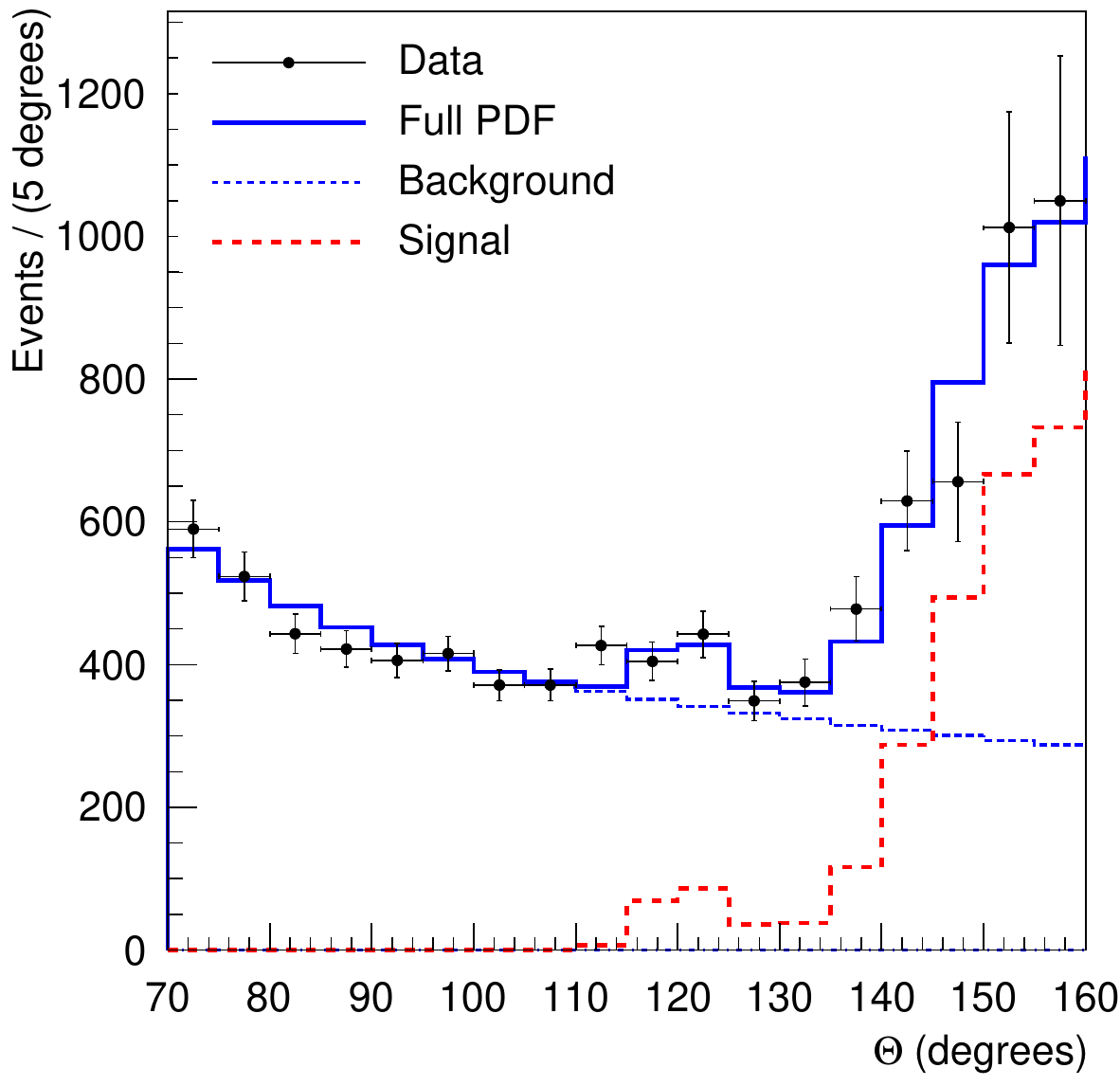}
    \end{center}
\caption{Experimental angular correlations of the e$^+$e$^-$ pairs
  fitted by the contributions from the E1 IPC and from the
  contributions coming from the e$^+$e$^-$ decay of the X17 particle.}
     \label{Fig:ang-fit}
\end{figure}

The experimental e$^+$e$^-$ angular correlation was fitted  with the following
intensity function
(INT) simulated as a function of the invariant mass:
\begin{equation}
\label{eq:pdf}
\begin{split}
INT&(e^+e^-) = \\
 &N_{E1} * PDF(E1) + N_{M1} * PDF(M1) + \\
 &N_{Sig} * \alpha_{ground}* PDF(sigground) + \\
 &N_{Sig} * (1 - \alpha_{ground}) * PDF(sig2plus)\ ,
\end{split}
\end{equation}

\noindent
where $PDF(X)$ represents the MC-simulated probability density
functions. $PDF(E1)$ and $PDF(M1)$ were simulated for Internal Pair
Creation having electromagnetic transitions with E1 and M1
multipolarity. $PDF(sigground)$ and $PDF(sig2plus)$ were simulated for the
two-body decay of an X17 particle as a function of its mass created in
the GDR to the ground state and GDR to $2_1^+$ transitions,
respectively.  $N_{E1}$, $N_{M1}$, and $N_{Sig}$ are the fitted
numbers of background and signal events, respectively.
$\alpha_{ground}$ is the fraction of X17 decays detected in the GDR to
ground state transition, with respect to the total number of detected
X17 decays. We assumed the same mass for the X17 particle created in
the two transitions.  The result of the fit is shown in
Fig.~\ref{Fig:ang-fit} together with the experimental data.

As shown in Fig.~\ref{Fig:ang-fit}, the simulation can describe the
experimental distributions from $\Theta = 70^\circ$ to 160$^\circ$
well. The fit results in a  $>10\sigma$ confidence for the X17
hypothesis.

The measured invariant mass of the hypothetical X17 particle was
determined as the value with which the above INT function provided the
best fit to the experimental data.
The obtained value for the invariant mass is
$m_\mathrm{X}c^2$~=~16.94~$\pm$~0.47(stat.)~MeV/c$^2$, which agrees
well with the invariant masses that we
obtained previously \cite{kr16,kr21,kr22}.

The systematic uncertainties were estimated to be
${\Delta}m_\mathrm{X}c^2$(syst.)$~=~\pm~0.35$~MeV by employing a series of MC
simulations as presented in one of our previous works \cite{kr21}. It
mostly represents the uncertainty of the position of the beam spot,
which was found to be shifted by about $\pm 3$~mm in one measurement
run.

The intensity ratio of the X17 particle emission to the ground state
($J^\pi = 0^+$) and to the first excited state $J^\pi = 2^+$ was found
to be:
\begin{equation}
\label{eq:alpha}
\begin{split}
\frac{B_{X17}(GDR \rightarrow g.s.)}{B_{X17}(GDR \rightarrow 2_1^+)} =
  \frac{\alpha_{ground}}{1 - \alpha_{ground}} =
  0.08 \pm 0.08 \ .
\end{split}
\end{equation}
\noindent

\section{Summary}

We reported on a new direction of the X17 research. For the first time, we
successfully detected this particle in the decay of a Giant Dipole
Resonance (GDR). Since this resonance is a general property of all
nuclei, the study of GDR may extend these studies to the entire
nuclear chart.

We have studied the GDR (J$^\pi$~=~1$^-$) E1-decay to the ground state
(J$^\pi$~=~$0^+$) and to the first excited state (J$^\pi$~=~$2_1^+$) in
$^{8}$Be.  The energy-sum and the angular correlation of the e$^+$e$^-$
pairs produced in the $^{7}$Li($p$,e$^+$e$^-$)$^{8}$Be reaction was
measured at a proton energy of E$_p$~=~4.0~MeV.  The gross features of
the angular correlation can be well-described by the IPC process
following the decay of the GDR.  However, on top of the smooth,
monotonic distribution of the angular correlation of the e$^+$e$^-$ pairs,
we observed significant anomalous excesses at about 120$^\circ$ and
above 140$^\circ$.

The e$^+$e$^-$ excesses can be well-described by the creation and
subsequent decay of the X17 particle, which we have recently suggested
\cite{kr16,kr21,kr22}. The invariant mass of the particle was measured
to be ($m_\mathrm{X}c^2 = 16.95 \pm 0.48$(stat.)$\pm 0.35$(syst.)~MeV),
which agrees well with our previous results.

The present observation of the X17 particle in  E1 transitions
supports its vector or axial vector character if it is emitted with an L=0 or L=1
angular momentum.

\section{Conclusion}

The spins and parities of the nuclear states examined in our
experiments are well known. For $^8$Be, for example, the 18.15~MeV
state has J$^\pi$~=~1$^+$ spin and parity. Based on this, we could
think that the X17 particle, which was created during the decay from
these states to the ground state, has spin and parity J$^\pi$~=~1$^+$,
i.e. a so-called axial-vector particle.

However, the anomaly was also observed in the 1$^-\rightarrow$ 0$^+$
transition of $^{12}$C, when a J$^\pi$~=~1$^-$ vector type particle
with L=0 orbital momentum could exit the nucleus. However, it is also
possible in this case to exit with L=1 orbital angular momentum. Then
J$^\pi$~=~1$^+$ axial vector particles could also leave the nucleus.

Currently, we have no experimental information on the orbital momentum
of the X17 particle in either case. Thus, we cannot say whether the
X17 particle is of a vector or an axial vector type.

The orbital angular momentum of the particles leaving the nucleus is
usually determined based on the angular distribution measured with
respect to the direction of the particle (proton) that created the
nuclear reaction. In the case of exit with L=0, this angular distribution
is expected to be isotropic, and in the case of L=1, it is expected to
be anisotropic. 

The currently running experiments having large acceptance
can provide information on the
angular momentum of the exiting X17 particle, so we eagerly
await their results.

\section{Acknowledgements}

We wish to thank Z. Pintye for the mechanical and J. Moln\'ar for the
electronic design of the experiment.  This work has been supported by
the GINOP-2.3.3-15-2016-00034 and
\noindent GINOP-2.3.3-15-2016-00005 grants. Partial support of
this talk, presented at the ISMD-2023 conference, by MVM NPP, Paks, Hungary,
https://atomeromu.mvm.hu/en/ is also greatfully acknowledged.

\end{document}